\documentclass[aps,prd,reprint,amsmath,amssymb,superscriptaddress,longbibliography]{revtex4-1}

\usepackage{graphicx}
\usepackage{bm}
\usepackage{braket}
\usepackage{hyperref}
\hypersetup{colorlinks=true,allcolors=blue}

\begin{document}

\title{Imaginary pseudo entropy encodes temporal orientation}

\author{Tatsuhiro Misumi}
\email{misumi@phys.kindai.ac.jp}
\affiliation{Department of Physics, Kindai University, Higashi-Osaka, Osaka 577-8502, Japan}

\begin{abstract}
Pseudo entropy between quantum states at different times is generally complex, yet its imaginary part has lacked a bounded operational meaning. We show that a calibrated replica interferometer converts the pseudo-R\'enyi phase into a directly measurable record of transition orientation. Together with replica visibility, it exactly determines the trace distance between forward and backward ancilla outputs and hence the Helstrom-optimal single-shot success probability. At short times, the symmetrized covariance of the modular and physical Hamiltonians sets the initial distinguishability response. Under any common quantum channel, the corresponding orientation information can only decrease, with equality characterized by Petz recovery. Imaginary pseudo entropy therefore records a reversible distinction between temporal orientations, while coarse graining can make the loss of that record irreversible.
\end{abstract}

\maketitle

\section{Introduction}
For a bipartite pure state, R\'enyi moments of the reduced density matrix, which underlie the R\'enyi entanglement entropies, can be measured by permuting identically prepared copies \cite{Renyi1961,Calabrese2004,Ekert2002,Abanin2012,Daley2012,Islam2015}. Pseudo entropy replaces the density matrix by a normalized transition matrix between two nonorthogonal states \cite{Nakata2021,Mollabashi2021} and is generally complex. Its complex structure has been studied in postselected, real-time, timelike, thermal, and spectral settings \cite{Mollabashi2021Aspects,Goto2021,Nishioka:2021cxe,Guo2022,Mukherjee:2022jac,Doi2023,Ishiyama2022,Narayan:2022afv,Narayan:2023ebn,Narayan:2023zen,Nanda:2025tid,Narayan:2026wzp,GuoReality2022,Chen2023,Caputa2025}. 
In holographic and timelike settings, the imaginary part has been connected to emergent time and twist-operator commutators \cite{Doi2023,Doi:2023zaf,GuoXu2025}.
Related spacetime-density-matrix approaches formulate temporal correlations between different Cauchy surfaces
and lead to generally non-Hermitian reduced operators for causally connected subsystems \cite{Milekhin2025,Guo:2025dtq},
while entanglement imagitivity provides a norm-based measure of temporal quantum correlations \cite{Milekhin2025}.
The spacetime-kernel approaches have also been developed to encode temporal correlations, time-ordered and out-of-time-ordered structures, and spectral-chaos diagnostics \cite{Das:2025fcd,Das:2026ifj}.
Our previous work showed that the short-time imaginary response of real-time pseudo entropy is the symmetrized covariance of the subsystem modular and physical Hamiltonians \cite{Misumi2026}.

What remained missing was a bounded operational meaning of the imaginary part itself. The issue is especially sharp near small overlaps, where normalization can strongly amplify pseudo entropy, much as in weak-value amplification \cite{Ishiyama2022,Aharonov1988}. Exchanging the initial and final states complex conjugates the reduced transition matrix and reverses the imaginary part \cite{Guo2022,Doi2023,Chen2023,Misumi2026}. We call this exchange a reversal of transition orientation. It need not coincide with antiunitary time reversal and does not by itself define a thermodynamic arrow: a closed unitary process may be reversible even when interference distinguishes its two orientations. The question is therefore how much information about this orientation can be read out and how much is lost under a specified coarse graining. Forward--backward distinguishability also appears in information-theoretic formulations of entropy production \cite{Kawai2007,Kwon2019,Landi2021}, but here the orientation record exists before dissipation, and its loss is not automatically thermodynamic entropy production.

We answer these questions with a phase-calibrated replica interferometer. The pseudo-R\'enyi phase and measured visibility exactly determine the trace distance between the forward and backward ancilla states and hence their Helstrom-optimal single-shot discrimination probability, attained by an ancilla $Y$ measurement. 
At finite replica number, the short-time signal is directly
accessible through the replica protocol; in the von Neumann limit, the modular covariance found in \cite{Misumi2026} becomes the normalized initial rate of optimal orientation distinguishability. 
Finally, quantum relative entropy measures the orientation information carried by the ancilla pair. Applying the same coarse graining to both states can only reduce this information, with exact preservation characterized by Petz recovery. The imaginary part thus provides a measurable record of temporal orientation; irreversibility enters only when coarse graining destroys part of that record.

\begin{figure*}[t]
 \includegraphics[width=\textwidth]{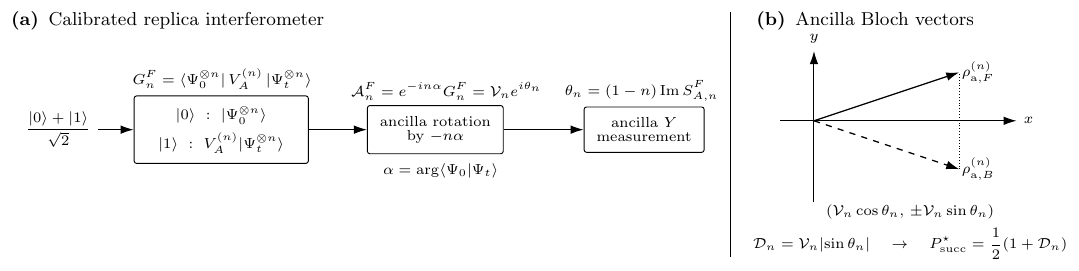}
 \caption{
 Calibrated replica interferometer and optimal orientation readout. (a) The ancilla controls the branches $\ket{\Psi_0^{\otimes n}}$ and $V_A^{(n)}\ket{\Psi_t^{\otimes n}}$, whose overlap is $G_n^F$. A one-copy calibration determines $\alpha=\arg\chi$; the rotation by $-n\alpha$ removes the normalization phase and leaves $\mathcal A_n^F=\mathcal V_n e^{i\theta_n}$. (b) Reversing the transition orientation sends $\theta_n\to-\theta_n$, giving opposite $y$ components. Their trace distance is $D_n=\mathcal V_n|\sin\theta_n|$, which yields $P_{\rm succ}^\star=(1+D_n)/2$; the ancilla $Y$ measurement is Helstrom optimal.
}
\label{fig:interferometer}
\end{figure*}

\section{Real-time pseudo-R\'enyi phase}
Let $\mathcal H=\mathcal H_A\otimes\mathcal H_{\bar A}$, with normalized initial state $|\Psi_0\rangle$ and unitary evolution $|\Psi_t\rangle=e^{-iHt}|\Psi_0\rangle$. Define the unnormalized forward reduced transition matrix and total overlap by
\begin{align}
 T_A^F(t)=\operatorname{Tr}_{\bar A}|\Psi_t\rangle\langle\Psi_0|,
 \qquad
 \chi(t)=\langle\Psi_0|\Psi_t\rangle.
 \label{eq:transition}
\end{align}
For $\chi(t)\neq0$, define $\tau_A^F(t)=T_A^F(t)/\chi(t)$. We suppress the argument $t$ below when no confusion can arise. For integer $n\geq2$, introduce
\begin{align}
 Z_n^F=\operatorname{Tr}_A[(\tau_A^F)^n]
 =\frac{G_n^F}{\chi^n},
 \qquad
 S_{A,n}^F=\frac{1}{1-n}\log Z_n^F,
 \label{eq:pseudoRenyi}
\end{align}
where $G_n^F=\operatorname{Tr}_A[(T_A^F)^n]$. We choose the logarithmic branch continuously from $t=0$ on any interval containing no zero of $Z_n^F$ or $\chi$. Physical results below depend on $\sin\arg Z_n^F$ and are therefore invariant under branch shifts by $2\pi$.

The opposite transition orientation is obtained by exchanging bra and ket:
$T_A^B=\operatorname{Tr}_{\bar A}|\Psi_0\rangle\langle\Psi_t|=(T_A^F)^\dagger$. Consequently, $Z_n^B=(Z_n^F)^*$ and
\begin{align}
 \theta_n\equiv\arg Z_n^F=(1-n)\operatorname{Im}S_{A,n}^F,
 \qquad
 \arg Z_n^B=-\theta_n.
 \label{eq:phase}
\end{align}
We call this exchange a reversal of transition orientation. Let $\Theta$ denote an antiunitary time-reversal operator. 
The exchange $F\leftrightarrow B$ only swaps the ket and bra. It represents a physical time-reversed protocol only when $\Theta H\Theta^{-1}=H$ and the preparation and readout are transformed consistently by $\Theta$. Otherwise, $F$ and $B$ denote only the two algebraic orientations of the same transition amplitude.
Keeping this distinction explicit prevents a nonzero imaginary pseudo entropy in a closed system from being mistaken for dissipative irreversibility. None of the state-discrimination results below requires an antiunitary symmetry.

\section{Replica interferometer and optimal discrimination}
\label{sec:Rep}
A replica is simply an independently prepared copy of the same bipartite system. We denote the full $n$-copy state by
$|\Psi^{\otimes n}\rangle\equiv|\Psi\rangle_1\otimes\cdots\otimes|\Psi\rangle_n$.
The unitary $V_A^{(n)}$ cyclically moves only the subsystem-$A$ factor of each copy to the next copy, while acting as the identity on all $\bar A$ factors. If $|a_k\rangle$ is a basis state of subsystem $A$ in copy $k$, then
\begin{align}
 V_A^{(n)}|a_1,a_2,\ldots,a_n\rangle_A
 =|a_n,a_1,\ldots,a_{n-1}\rangle_A.
 \label{eq:cyclic-action}
\end{align}
With this convention, contraction of the copy indices gives
\begin{align}
 G_n^F=\langle\Psi_0^{\otimes n}|V_A^{(n)}|\Psi_t^{\otimes n}\rangle.
 \label{eq:replicaoverlap}
\end{align}
This is the transition-matrix analogue of the permutation identities used to measure state moments and R\'enyi entropies \cite{Ekert2002,Abanin2012,Daley2012,Islam2015}; the index contraction is shown explicitly in Appendix \ref{app:1}. Because Eq.~\eqref{eq:replicaoverlap} is an overlap between normalized many-copy states, $|G_n^F|\leq1$. It can therefore appear as the off-diagonal coherence of an auxiliary two-level control system, which we call the ancilla qubit. The normalized quantity $Z_n^F$, by contrast, can have modulus larger than one and cannot in general be used directly as a physical coherence.

Prepare the ancilla in $(|0\rangle+|1\rangle)/\sqrt2$. In the $|0\rangle$ branch, prepare $|\Psi_0^{\otimes n}\rangle$; in the $|1\rangle$ branch, prepare $V_A^{(n)}|\Psi_t^{\otimes n}\rangle$. Their overlap is $G_n^F$. Writing $\chi=|\chi|e^{i\alpha}$, a separate one-copy interference measurement determines $\alpha$. A compensating ancilla rotation by $-n\alpha$ then removes the phase coming solely from the normalization $\chi^n$ and leaves the calibrated amplitude
\begin{align}
 \mathcal A_n^F=e^{-in\alpha}G_n^F
 =\mathcal V_n e^{i\theta_n},
 \qquad
 \mathcal V_n \equiv|G_n^F|.
 \label{eq:calibrated}
\end{align}
A fully coherent reference-register implementation is described in Appendix \ref{app:1}. The backward amplitude is $(\mathcal A_n^F)^*$. Tracing out the replica registers gives the forward and backward ancilla density matrices,
\begin{align}
 \rho_{\mathrm a,F}^{(n)}&=\frac12\left[I+\mathcal V_n\cos\theta_n\,\sigma_x
 +\mathcal V_n\sin\theta_n\,\sigma_y\right],\nonumber\\
 \rho_{\mathrm a,B}^{(n)}&=\frac12\left[I+\mathcal V_n\cos\theta_n\,\sigma_x
 -\mathcal V_n\sin\theta_n\,\sigma_y\right].
 \label{eq:ancillastates}
\end{align}
Their common $\sigma_x$ component carries orientation-even information, while their opposite $\sigma_y$ components contain all orientation-odd information accessible in the calibrated interferometer. The trace distance is the standard quantum measure of how well two states can be distinguished: it is zero for identical states and one for perfectly distinguishable states. Here $\|X\|_1=\operatorname{Tr}\sqrt{X^\dagger X}$ denotes the trace norm, and
\begin{align}
 \mathcal D_n
 =\frac12\|\rho_{\mathrm a,F}^{(n)}-\rho_{\mathrm a,B}^{(n)}\|_1
 =\mathcal V_n\left|\sin\!\left[(1-n)\operatorname{Im}S_{A,n}^F\right]\right|.
 \label{eq:mainresult}
\end{align}
Note $(1-n)\operatorname{Im}S_{A,n}^F =\theta_n$ and the details of calculations are given in Appendix \ref{app:2}.
Eq.~\eqref{eq:mainresult} and Fig.~\ref{fig:interferometer} are our central results. The imaginary pseudo-R\'enyi entropy fixes the opposite azimuthal angles $\pm\theta_n$ of the two output Bloch vectors, whereas $\mathcal V_n$ measures the coherence available to reveal them. Both quantities are obtained from the same interference fringe: $\mathcal V_n$ is its contrast and $\theta_n$ is its calibrated phase. For integer $n$, testing Eq.~\eqref{eq:mainresult} requires no analytic continuation and no reconstruction of the non-Hermitian operator $T_A^F$. The identity holds for an arbitrary bipartite state and unitary evolution; it does not rely on conformal symmetry, holography, or a semiclassical limit. Thus a large phase is not useful when the visibility vanishes, and a high-visibility interferometer is orientation blind when the pseudo-R\'enyi phase vanishes. The visibility is essential near a zero of $G_n^F$: its phase may wind rapidly, but $\mathcal V_n\to0$ suppresses the distinguishability. A zero of $\chi$ is different because the normalized transition matrix itself is then undefined. Our result applies on each connected time interval with $\chi\neq0$ and is insensitive there to the chosen continuous branch of the pseudo entropy.

The distinction between the normalized quantity $Z_n^F$ and the physical coherence $G_n^F$ also resolves the apparent tension between pseudo-entropy amplification and probability bounds.
The factor $\chi^{-n}$ can make $Z_n^F$ and $S_{A,n}^F$ large when the two states have a small overlap, much as in weak-value amplification. The physical ancilla coherence, however, is $G_n^F$, not $Z_n^F$. Consequently $0\leq\mathcal V_n\leq1$ and $0\leq\mathcal D_n\leq1$ even when the normalized pseudo entropy is parametrically enhanced. The visibility-dressed combination in Eq.~\eqref{eq:mainresult}, rather than the phase alone, is the bounded operational quantity.

For equal prior probabilities, the Holevo--Helstrom theorem gives the optimal success probability \cite{Holevo1973,Helstrom1976}
\begin{align}
 P_{\mathrm{succ}}^\star=\frac12(1+\mathcal D_n).
 \label{eq:helstrom}
\end{align}
Since $\rho_{\mathrm a,F}^{(n)}-\rho_{\mathrm a,B}^{(n)} \propto\sigma_y$, the Helstrom measurement is simply the ancilla $Y$ measurement; no further optimization of the readout basis is required.
Fig.~\ref{fig:interferometer} summarizes both the circuit and the Bloch-sphere geometry. For the optimal ancilla $Y$ measurement, let $y=\pm1$ denote the measured eigenvalue. The corresponding outcome probabilities for the forward and backward states are $p_F(y)=(1+yx_n)/2,\, p_B(y)=(1-yx_n)/2$ where
\begin{align}
 x_n=\mathcal V_n\sin\theta_n,
 \qquad \mathcal D_n=|x_n|.
 \label{eq:signedsignal}
\end{align}
The sign of $x_n$ records the orientation, while its magnitude determines the single-shot advantage, $P_{\mathrm{succ}}^\star-1/2=|x_n|/2$. The optimality here refers to discrimination of the two calibrated ancilla outputs produced by the replica protocol. Within that output space no other positive-operator-valued measurement can perform better. The protocol does not require tomography of $T_A^F$ or diagonalization of a non-Hermitian matrix. It requires $n$ coherent copies, a controlled subsystem permutation, a one-copy reference measurement of $\arg\chi$, and a final one-qubit readout. In particular, $n=2$ reduces the permutation to a controlled SWAP and uses the same two-copy architecture as purity measurements \cite{Ekert2002,Abanin2012,Daley2012,Islam2015}. Loss of branch coherence is reflected directly in the measured $\mathcal V_n$, rather than being hidden in an assumed ideal contrast. A systematic reference-phase offset rotates the optimal readout axis and can be removed by the same $n=1$ calibration used to determine $\alpha$.
The relative entropy of the optimal outcomes and its small-phase expansion are given in Appendix \ref{app:2}.

\section{Microscopic generation of the temporal signal}
We now restore the time arguments as $\theta_n(t)=\arg Z_n^F(t)$, $x_n(t)=\mathcal V_n(t)\sin\theta_n(t)$.
Let $\rho_A=\operatorname{Tr}_{\bar A}|\Psi_0\rangle\langle\Psi_0|$, $P_n=\operatorname{Tr}_A\rho_A^n$ be its $n$th moment, and $\langle O\rangle_0=\langle\Psi_0|O|\Psi_0\rangle$. Direct differentiation gives
\begin{align}
 \dot\theta_n(0)
 =-n\left[
 \frac{\frac12\langle\{\rho_A^{n-1},H\}\rangle_0}{P_n}
 -\langle H\rangle_0\right],
 \label{eq:finite-n-rate}
\end{align}
where an operator on $A$ is understood to be tensored with the identity on $\bar A$. Equivalently, the directly measurable signed signal satisfies (see App.~\ref{app:3} for details)
\begin{align}
 \dot x_n(0)=-n\left[
 \frac12\langle\{\rho_A^{n-1},H\}\rangle_0
 -P_n\langle H\rangle_0\right].
 \label{eq:finite-n-xrate}
\end{align}
At finite integer $n$, Eq.~\eqref{eq:finite-n-xrate} is a replica covariance and can be accessed without reconstructing the modular Hamiltonian. For $n=2$, let $\mathbb S_A\equiv V_A^{(2)}$ denote the subsystem SWAP operator between the two copies and let $H_\Sigma=H^{(1)}+H^{(2)}$. Here $\Delta O=O-\langle O\rangle_{\Psi_0^{\otimes2}}$. Then
\begin{align}
 \dot x_2(0)=-\frac12\left\langle\left\{\Delta\mathbb S_A,\Delta H_\Sigma\right\}\right\rangle_{\Psi_0^{\otimes2}}.
 \label{eq:swapcovariance}
\end{align}
The initial signal is therefore a symmetrized covariance between a subsystem SWAP and the two-copy Hamiltonian, without assumptions of integrability, Gaussianity, or Schmidt-diagonal form. 
For a weak quench $H_0\to H_0+\delta\lambda {\mathcal O}$ from an eigenstate of $H_0$, Eq.~\eqref{eq:swapcovariance} reduces to a linear susceptibility to $\mathcal O$; see Appendices \ref{app:3} and \ref{app:7}.
Ref.~\cite{Misumi2026} showed that the corresponding von Neumann modular susceptibility peaks near the Ising critical region. Here we compute the finite-replica phase response and the visibility-dressed discrimination response in the open transverse-field Ising chain. Both responses develop finite-size peaks whose positions approach the critical field, while their amplitudes differ by the replica visibility. 
At the critical field, boundary conformal field theory predicts that the half-chain visibility scales as $P_2\propto L^{-1/16}$. Our numerical data up to $L=1024$ show that the phase response remains of order unity, so the visibility-dressed discrimination response is only weakly reduced over the sizes studied.
The calculation uses the exact Ising-chain solution and Gaussian correlation-matrix methods~\cite{Pfeuty1970,Peschel2003,PeschelEisler2009,IgloiLin2008}; numerical details and results are given in Appendix \ref{app:7}.

Assuming a smooth continuation to $n=1$ and writing $K_A=-\log\rho_A$, Eq.~\eqref{eq:finite-n-xrate} reproduces the modular-covariance relation derived in \cite{Misumi2026}:
\begin{align}
 \left.\frac{d}{dt}\operatorname{Im}S_A^F(t)\right|_{t=0}
 =-\frac12\langle\{\Delta K_A,\Delta H\}\rangle_0
 \equiv-\mathcal C_{KH}.
 \label{eq:modularcovariance}
\end{align}
Combining Eqs.~\eqref{eq:mainresult} and \eqref{eq:modularcovariance} yields
\begin{align}
 \lim_{n\to1}\lim_{t\to0}
 \frac{\mathcal D_n(t)}{|n-1|\,|t|}=|\mathcal C_{KH}|.
 \label{eq:rate-dist}
\end{align}
Thus the covariance derived in \cite{Misumi2026} is the normalized initial rate at which optimal temporal-orientation distinguishability is generated. Integer $n$ is directly accessible interferometrically, whereas the $n\to1$ limit identifies the von Neumann pseudo-entropy slope. The vanishing of $\mathcal D_n$ as $n\to1$ therefore reflects the kinematic factor $|n-1|$, rather than a disappearance of the underlying response. 
For imaginary-time R\'enyi entropy $S_{A,n}^{E}(\tau)$, $\left.\partial_t\operatorname{Im}S_{A,n}^{F}\right|_{0}=\frac12\left.\partial_\tau S_{A,n}^{E}\right|_{0}$ is useful for many-body calculations; see Appendix \ref{app:3}.


\begin{figure*}[t]
 \includegraphics[width=0.9\textwidth]{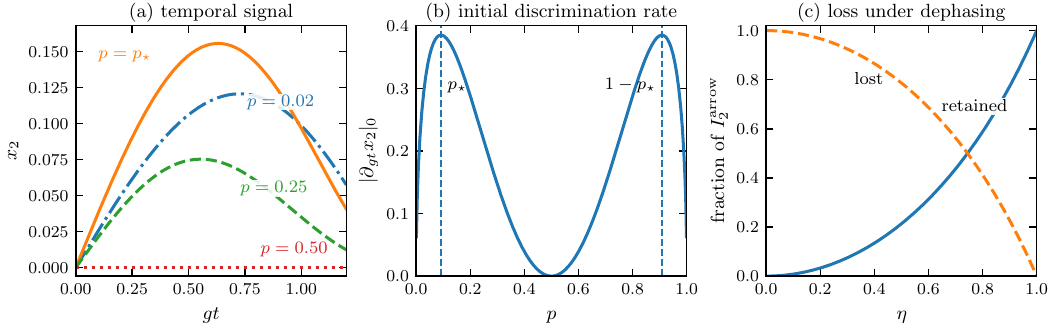}
 \caption{
 Two-qubit realization for a real initial state. (a) Signed temporal signal $x_2$ for representative Schmidt weights, with $p_\star=(1-\sqrt{2/3})/2$. (b) The initial discrimination rate $|\partial_{gt}x_2|_0=2\sqrt{p(1-p)}(2p-1)^2$ is largest at $p=p_\star$ and $1-p_\star$, rather than at a product state or a maximally entangled state. (c) For $p=p_\star$ and $gt=0.45$, dephasing continuously transfers the initially available arrow information from the retained to the lost fraction.}
 \label{fig:twoqubit}
\end{figure*}

\section{Loss of orientation information under coarse graining}
$\mathcal D_n$ in Eq.~\eqref{eq:mainresult} concerns coherent temporal orientation and can be nonzero even for an exactly reversible unitary evolution. To quantify asymptotic discrimination of the ancilla pair, define
\begin{align}
 I_n^{\mathrm{arrow}}
 =D(\rho_{\mathrm a,F}^{(n)}\|\rho_{\mathrm a,B}^{(n)}),
 \label{eq:arrowinfo}
\end{align}
where $D(\rho\|\sigma)=\operatorname{Tr}\rho(\log\rho-\log\sigma)$ is the quantum relative entropy \cite{Umegaki1962}. By the quantum Stein lemma, it gives the optimal asymmetric many-copy error exponent \cite{HiaiPetz1991,OgawaNagaoka2000}, complementing the single-shot trace distance. For the mirror-related states in Eq.~\eqref{eq:ancillastates},
\begin{align}
 I_n^{\mathrm{arrow}}
 =2\mathcal V_n\sin^2\theta_n\operatorname{artanh}\mathcal V_n,
 \qquad 0\leq\mathcal V_n<1.
 \label{eq:arrowinfo-explicit}
\end{align}
Thus the same visibility and phase determine both single-shot and asymptotic distinguishability. Operationally, $\mathcal D_n$ governs the best one-shot decision, whereas $I_n^{\mathrm{arrow}}$ governs the asymptotic error exponent. 
The classical relative entropy of the optimal $Y$-outcome distributions is $D_{\mathrm{KL}}(p_F\Vert p_B)=2\mathcal D_n\operatorname{artanh}\mathcal D_n$ and satisfies $I_n^{\mathrm{arrow}}\geq D_{\mathrm{KL}}\geq2\mathcal D_n^2$; see Appendix \ref{app:6}.

A nonzero $I_n^{\mathrm{arrow}}$ is compatible with closed, reversible dynamics: unitary processing preserves relative entropy even when the two orientations remain distinguishable. The pseudo-entropy phase is therefore a coherent record, not dissipative entropy production.

Let the same completely positive trace-preserving map $\Lambda$ act on both outputs. The common map may describe detector noise, dephasing, discarding an environment, finite resolution, or conversion to a classical record; using different maps could itself create distinguishability. Monotonicity of quantum relative entropy \cite{Lindblad1975,Uhlmann1977} gives
\begin{align}
 \Delta_\Lambda I_n^{\mathrm{arrow}}
 \equiv I_n^{\mathrm{arrow}}
 -D\!\left(\Lambda(\rho_{\mathrm a,F}^{(n)})\middle\|\Lambda(\rho_{\mathrm a,B}^{(n)})\right)
 \geq0.
 \label{eq:arrowloss}
\end{align}
We interpret $\Delta_\Lambda I_n^{\mathrm{arrow}}$ as temporal-orientation information discarded by the chosen coarse graining. It is a relative quantity for the pair of protocols, not the entropy of either state. Relating it to heat or work additionally requires a physical open-system channel, a reference state, and specified reverse dynamics \cite{Kawai2007,Kwon2019,Landi2021}.

Under the usual support conditions, equality in
Eq.~\eqref{eq:arrowloss} holds if and only if the Petz
recovery map reconstructs both ancilla states from their
coarse-grained outputs~\cite{Petz1986}.
The explicit map is given in Appendix \ref{app:6}. Thus zero loss means that the coarse graining retains all information needed to reconstruct the pair, whereas positive loss quantifies an unrecoverable part of the orientation record. This criterion is stronger than preserving the trace distance alone because it requires recovery of the full pair. 
For the phase-damping channel
$\Lambda_\eta(\sigma_{x,y})=\eta\sigma_{x,y}$,
$0\leq\eta\leq1$, Eq.~\eqref{eq:arrowinfo} with
$\mathcal V_n\to\eta\mathcal V_n$ gives the retained information, while the loss is
\begin{align}
 \Delta_\eta 
 I_n^{\mathrm{arrow}}
 =2\mathcal V_n\sin^2\theta_n
 \left[\operatorname{artanh}\mathcal V_n-\eta\operatorname{artanh}(\eta\mathcal V_n)\right].
 \label{eq:dephasing}
\end{align}
The loss grows from zero at $\eta=1$ to complete removal of the phase-sensitive record at $\eta=0$. Thus unitary dynamics first creates orientation information and dephasing subsequently erases it. A thermodynamic interpretation would require additional structure---an actual reverse channel, a stationary or thermal reference state, and a relation to environmental entropy flow---none of which enters the present theorem.

\section{Two-qubit realization}
To illustrate the new operational quantities, we apply our replica protocol to the noncommuting two-qubit model introduced in \cite{Misumi2026},
$|\Psi_0\rangle=\sqrt p|00\rangle+\sqrt{1-p}|11\rangle$ and
$H=J\sigma_z\otimes\sigma_z+g\sigma_x\otimes\sigma_x$. For $n=2$, writing $r=\sqrt{p(1-p)}$, $c=\cos(gt)$, $s=\sin(gt)$, and $\delta=2p-1$, the calibrated signed signal is
\begin{align}
 x_2(t)=\frac{2r\delta^2c^3s}{1-\delta^2s^2},
 \qquad \mathcal D_2=|x_2|.
 \label{eq:twoqubit}
\end{align}
The coupling $J$ contributes only a common phase in the occupied $|00\rangle,|11\rangle$ sector, whereas $g$ mixes the Schmidt sectors and generates the signal. The signal vanishes for a product state, where $r=0$ and no coherent superposition of the two Schmidt sectors is available, and for a maximally entangled state, where the modular spectrum is flat and $\delta=0$. At short times, $x_2=2gr\delta^2t+O(t^3)$; its magnitude is maximal at $\delta^2=2/3$, or $p=(1\pm\sqrt{2/3})/2$. Temporal-orientation sensitivity therefore requires both coherence between Schmidt sectors and a nontrivial modular spectrum, and is optimized at intermediate entanglement. Fig.~\ref{fig:twoqubit} shows the signal, its optimal initial rate, and the subsequent information loss under dephasing. The Schmidt-diagonal class and complete two-qubit formulas are given in Appendices \ref{app:4} and \ref{app:5}.

\section{Conclusion and Discussion}
We have given the imaginary part of real-time pseudo entropy a bounded operational meaning. In a calibrated replica interferometer, the pseudo-R\'enyi phase and visibility determine the Helstrom-optimal discrimination of opposite transition orientations. The finite-replica response is a directly measurable covariance, while its von Neumann limit turns the modular covariance of \cite{Misumi2026} into the normalized initial distinguishability rate. 
Quantum relative entropy measures the distinguishability between the two orientations after coarse graining; when this distinguishability is exactly preserved, the original states can be recovered by the Petz map.
Unitary dynamics can create the orientation record, replica interference can read it out, and a noninvertible channel can erase it. Imaginary pseudo entropy is therefore not thermodynamic entropy production; it is a coherent orientation record whose loss becomes irreversible once a physical coarse graining is specified.

The proposal gives three direct experimental checks: reversing the orientation flips the calibrated ancilla $Y$ signal while leaving its $X$ component unchanged, the trace distance obeys Eq.~\eqref{eq:mainresult}, and the information loss under controlled dephasing follows Eq.~\eqref{eq:dephasing}. 
Because the protocol uses subsystem permutations rather than tomography of a non-Hermitian operator, it applies beyond integrable systems. Detailed derivations, complete two-qubit formulas, and the Ising-chain analysis are given in the appendices. 
Rapid winding of the calibrated phase as Fisher zeros approach the real-time axis may provide an interferometric signature of dynamical quantum phase transitions \cite{Heyl2013,Heyl2018}.




\begin{acknowledgements}
The author thanks Genki Tanaka for fruitful discussions related to his undergraduate thesis at Kindai University, from which this work grew. The author also thanks the organizers of the workshop ``Complexification 2024'' at Saga University. This work was supported by JSPS KAKENHI Grants No.~23K03425 and No.~22H05118.
\end{acknowledgements}

\appendix

\section{Replica-permutation identity and calibrated overlap}
\label{app:1}

These appendices provide detailed derivations, analytic formulas, and numerical analyses supporting the main text. We use the same notation throughout: $F$ and $B$ label the two transition orientations, $\rho_A$ is the initial reduced density matrix, $T_A^{F,B}$ are unnormalized reduced transition matrices, $\tau_A^{F,B}$ are their normalized counterparts, and $\rho_{\mathrm a,F/B}^{(n)}$ are the ancilla outputs. All expectation values $\langle\cdots\rangle_0$ are evaluated in the initial state $|\Psi_0\rangle$.

For normalized states $|\psi\rangle,|\phi\rangle\in\mathcal H_A\otimes\mathcal H_{\bar A}$, define
\begin{align}
 T_A=\operatorname{Tr}_{\bar A}|\psi\rangle\langle\phi|.
\end{align}
We write $|\psi^{\otimes n}\rangle=|\psi\rangle_1\otimes\cdots\otimes|\psi\rangle_n$ for $n$ independently prepared copies. The unitary $V_A^{(n)}$ cyclically permutes their $A$ factors according to $V_A^{(n)}|a_1,\ldots,a_n\rangle_A=|a_n,a_1,\ldots,a_{n-1}\rangle_A$ and acts as the identity on all $\bar A$ factors. In bases $|a\rangle_A$ and $|\mu\rangle_{\bar A}$,
\begin{align}
 \langle a|T_A|b\rangle
 =\sum_\mu\langle a,\mu|\psi\rangle
 \langle\phi|b,\mu\rangle.
\end{align}
Multiplying $n$ matrix elements and tracing the $A$ indices yields
\begin{align}
 \operatorname{Tr}_{A}(T_A^n)
 =\langle\phi^{\otimes n}|V_A^{(n)}|\psi^{\otimes n}\rangle,
 \label{eq:SM-replica}
\end{align}
for the cyclic convention stated above. This is the transition-matrix version of the permutation identities used in direct measurements of state moments and R\'enyi entropies \cite{Ekert2002,Abanin2012,Daley2012,Islam2015}.

For the real-time transition, set $|\psi\rangle=|\Psi_t\rangle$ and $|\phi\rangle=|\Psi_0\rangle$. Define the forward and opposite-orientation reduced transition matrices by
\begin{align*}
 T_A^F(t)&\equiv\operatorname{Tr}_{\bar A}|\Psi_t\rangle\langle\Psi_0|,\\
 T_A^B(t)&\equiv\operatorname{Tr}_{\bar A}|\Psi_0\rangle\langle\Psi_t|
 =[T_A^F(t)]^\dagger.
\end{align*}
The total overlap is
\begin{align}
 \chi(t)\equiv\langle\Psi_0|\Psi_t\rangle
 =|\chi(t)|e^{i\alpha(t)}.
\end{align}
For $\chi(t)\neq0$, define
\begin{align*}
 \tau_A^F(t)&\equiv\frac{T_A^F(t)}{\chi(t)},
 &
 \tau_A^B(t)&\equiv\frac{T_A^B(t)}{\chi(t)^*},\\
 G_n^F(t)&\equiv\operatorname{Tr}_A[(T_A^F(t))^n],
 &
 G_n^B(t)&\equiv\operatorname{Tr}_A[(T_A^B(t))^n],\\
 Z_n^F(t)&\equiv\operatorname{Tr}_A[(\tau_A^F(t))^n],
 &
 Z_n^B(t)&\equiv\operatorname{Tr}_A[(\tau_A^B(t))^n],
\end{align*}
and
\begin{equation*}
 S_{A,n}^{F,B}(t)\equiv\frac{1}{1-n}\log Z_n^{F,B}(t).
\end{equation*}
We choose the logarithmic branch continuously from $t=0$ on any interval containing no zero of $\chi(t)$ or $Z_n^F(t)$. Equivalently, $\theta_n(t)$ below is defined away from zeros of $G_n^F(t)$; at such a zero, $\mathcal V_n(t)$ and the corresponding distinguishability vanish continuously.
In particular,
\begin{equation*}
 Z_n^F(t)=\frac{G_n^F(t)}{\chi(t)^n},
 \quad
 G_n^B(t)=[G_n^F(t)]^*,
 \quad
 Z_n^B(t)=[Z_n^F(t)]^*.
\end{equation*}
A compensating ancilla phase rotation by $-n\alpha(t)$ multiplies the $|1\rangle$ branch relative to the $|0\rangle$ branch by $e^{-in\alpha(t)}$. The phase-calibrated overlap is
\begin{align}
 \mathcal A_n^F(t)=e^{-in\alpha(t)}G_n^F(t)
 =\mathcal V_n(t)e^{i\theta_n(t)},
 \label{eq:SM-calibrated}
\end{align}
where
\begin{align}
 \mathcal V_n(t)&=|G_n^F(t)|,
 \nonumber\\
 \theta_n(t)
 &=\arg G_n^F(t)-n\arg\chi(t)
 \nonumber\\
 &=\arg Z_n^F(t)=(1-n)\operatorname{Im}S_{A,n}^F(t).
\end{align}
A fully coherent realization of the phase calibration may instead use an additional reference register whose branch overlap is $[\chi(t)^*]^n$. The resulting total branch coherence is
\begin{equation*}
 G_n^F(t)[\chi(t)^*]^n
 =|\chi(t)|^n\mathcal V_n(t)e^{i\theta_n(t)}.
\end{equation*}
Thus this realization yields the same calibrated phase $\theta_n(t)$, while its raw fringe visibility contains the additional factor $|\chi(t)|^n$. The ancilla-state formulas below refer to the externally calibrated protocol of the main text, for which the visibility is $\mathcal V_n(t)=|G_n^F(t)|$. If the applied relative phase differs from $-n\alpha(t)$ by a residual offset $\delta_{\rm cal}$, the measured Bloch vector is rotated by $\delta_{\rm cal}$ in the $xy$ plane. The same one-copy reference interferometer determines this offset, so the calibrated formulas correspond to $\delta_{\rm cal}=0$.

\section{Ancilla states and optimal orientation discrimination}
\label{app:2}

We suppress the common time argument $t$ throughout this section. Using one ancilla qubit as the branch control, prepare $|u\rangle=|\Psi_0^{\otimes n}\rangle$ in the $|0\rangle$ branch and $|v_F\rangle=e^{-in\alpha}V_A^{(n)}|\Psi_t^{\otimes n}\rangle$ in the $|1\rangle$ branch. The joint state is
\begin{align}
 |\Omega_F\rangle
 =\frac{1}{\sqrt2}\left(|0\rangle|u\rangle
 +|1\rangle|v_F\rangle\right),
 \quad
 \langle u|v_F\rangle=\mathcal V_ne^{i\theta_n}.
\end{align}
Tracing out the replica registers gives the forward and backward ancilla density matrices,
\begin{align}
 \rho_{\mathrm a,F}^{(n)}
 =\frac12\begin{pmatrix}
 1&\mathcal V_ne^{-i\theta_n}\\
 \mathcal V_ne^{i\theta_n}&1
 \end{pmatrix}.
 \label{eq:SM-rhoF}
\end{align}
The opposite transition orientation gives $\rho_{\mathrm a,B}^{(n)}=(\rho_{\mathrm a,F}^{(n)})^*$ and hence
\begin{align}
 \rho_{\mathrm a,F}^{(n)}-\rho_{\mathrm a,B}^{(n)}
 =\mathcal V_n\sin\theta_n\,\sigma_y.
\end{align}
Its eigenvalues are $\pm\mathcal V_n\sin\theta_n$, so the trace distance is
\begin{align}
 \mathcal D_n
 =\frac12\|\rho_{\mathrm a,F}^{(n)}-\rho_{\mathrm a,B}^{(n)}\|_1
 =\mathcal V_n|\sin\theta_n|.
\end{align}
The projectors onto the positive and negative eigenspaces are the $Y$-basis projectors. The Holevo-Helstrom theorem therefore gives \cite{Holevo1973,Helstrom1976}
\begin{align}
 P_{\mathrm{succ}}^\star=\frac12(1+\mathcal D_n).
\end{align}

For the optimal ancilla $Y$ measurement, let $y=\pm1$ denote the measured eigenvalue and let
\begin{equation*}
 \Pi_y\equiv\frac{\mathbb I+y\sigma_y}{2}
\end{equation*}
be the corresponding projector. The outcome probabilities for the two transition orientations are
\begin{align}
 &p_F(y)\equiv\operatorname{Tr}[\Pi_y\rho_{\mathrm a,F}^{(n)}]
 =\frac{1+yx_n}{2},
\nonumber\\
 &p_B(y)\equiv\operatorname{Tr}[\Pi_y\rho_{\mathrm a,B}^{(n)}]
 =\frac{1-yx_n}{2},
\nonumber\\
 &x_n\equiv\mathcal V_n\sin\theta_n.
\end{align}
Their Kullback--Leibler divergence, or classical relative entropy, is defined by
\begin{equation*}
 D_{\mathrm{KL}}(p_F\|p_B)
 \equiv\sum_{y=\pm1}p_F(y)\log\frac{p_F(y)}{p_B(y)}.
\end{equation*}
Using the probabilities above gives
\begin{align}
 D_{\mathrm{KL}}(p_F\|p_B)
 =x_n\log\frac{1+x_n}{1-x_n}
 =2\mathcal D_n\operatorname{artanh}\mathcal D_n.
\end{align}
For $|\theta_n|\ll 1$, this becomes
\begin{align}
D_{\rm KL}(p_F\|p_B)
&=2V_n^2\theta_n^2+O(\theta_n^4) \notag\\
&=2V_n^2(1-n)^2
\left(\operatorname{Im}S^F_{A,n}\right)^2
+O(\theta_n^4).
\end{align}

\section{Short-time response and imaginary-time identity}
\label{app:3}

At $t=0$, with $\rho_A\equiv\operatorname{Tr}_{\bar A}|\Psi_0\rangle\langle\Psi_0|$,
\begin{align}
 T_A^F(0)=\rho_A,
 \qquad
 G_n^F(0)=P_n\equiv\operatorname{Tr}_A\rho_A^n.
\end{align}
Furthermore,
\begin{align}
 \dot T_A^F(0)
 =-i\operatorname{Tr}_{\bar A}\left(H|\Psi_0\rangle\langle\Psi_0|\right).
\end{align}
Differentiating the moment and using cyclicity of the $A$ trace gives
\begin{align}
 \dot G_n^F(0)
 =-in\langle\rho_A^{n-1}H\rangle_0,
 \label{eq:SM-Gdot}
\end{align}
where $\rho_A^{n-1}$ is understood as $\rho_A^{n-1}\otimes\mathbb I_{\bar A}$. Since $G_n^F(0)=P_n$ is real and positive,
\begin{align}
 \left.\partial_t\arg G_n^F\right|_0
 =-\frac{n}{2P_n}\langle\{\rho_A^{n-1},H\}\rangle_0.
\end{align}
Using $\dot\chi(0)=-i\langle H\rangle_0$ therefore yields
\begin{align}
 \dot\theta_n(0)
 =-n\left[
 \frac{\frac12\langle\{\rho_A^{n-1},H\}\rangle_0}{P_n}
 -\langle H\rangle_0\right].
 \label{eq:SM-thetadot}
\end{align}
Because $x_n=\mathcal V_n\sin\theta_n$, $\theta_n(0)=0$, and $\mathcal V_n(0)=P_n$,
\begin{align}
 \dot x_n(0)
 =-n\left[
 \frac12\langle\{\rho_A^{n-1},H\}\rangle_0
 -P_n\langle H\rangle_0\right].
 \label{eq:SM-xdot}
\end{align}

Set $n=1+\epsilon$ and introduce the modular Hamiltonian $K_A=-\log\rho_A$ on the support of $\rho_A$. In full-system expectation values, $K_A$ is understood as $K_A\otimes\mathbb I_{\bar A}$. For any operator $O$, define $\Delta O\equiv O-\langle O\rangle_0$. The expansions
\begin{align}
 \rho_A^\epsilon=I-\epsilon K_A+O(\epsilon^2),
 \quad
 P_{1+\epsilon}=1-\epsilon\langle K_A\rangle_0+O(\epsilon^2)
\end{align}
give
\begin{align}
 \dot\theta_{1+\epsilon}(0)
 =\frac{\epsilon}{2}\langle\{\Delta K_A,\Delta H\}\rangle_0
 +O(\epsilon^2).
\end{align}
Assuming a smooth continuation to $n=1$, define the von Neumann pseudo entropy by $S_A^F(t)\equiv\lim_{n\to1}S_{A,n}^F(t)$. Since $\theta_n=(1-n)\operatorname{Im}S_{A,n}^F$, one obtains the modular-covariance relation derived in Ref.~\cite{Misumi2026},
\begin{align}
 \left.\partial_t\operatorname{Im}S_A^F\right|_0
 =-\frac12\langle\{\Delta K_A,\Delta H\}\rangle_0.
\end{align}

For normalized imaginary-time evolution
\begin{align}
 |\Psi_\tau\rangle
 =\frac{e^{-\tau H}|\Psi_0\rangle}
 {\sqrt{\langle\Psi_0|e^{-2\tau H}|\Psi_0\rangle}},
\end{align}
define
\begin{equation*}
 \rho_A(\tau)\equiv\operatorname{Tr}_{\bar A}|\Psi_\tau\rangle\langle\Psi_\tau|,
 \qquad
 P_n(\tau)\equiv\operatorname{Tr}_A[\rho_A(\tau)^n].
\end{equation*}
Then
\begin{align}
 \left.\partial_\tau|\Psi_\tau\rangle\right|_0
 =-(H-\langle H\rangle_0)|\Psi_0\rangle,
\end{align}
and
\begin{align}
 \left.\partial_\tau P_n(\tau)\right|_0
 =-n\langle\{\rho_A^{n-1},H-\langle H\rangle_0\}\rangle_0.
\end{align}
Defining $S_{A,n}^{\mathrm E}(\tau)\equiv(1-n)^{-1}\log P_n(\tau)$ and combining the preceding result with Eq.~\eqref{eq:SM-thetadot} gives
\begin{align}
 \left.\partial_t\operatorname{Im}S_{A,n}^F\right|_0
 =\frac12\left.\partial_\tau S_{A,n}^{\mathrm E}(\tau)\right|_0.
 \label{eq:SM-imagtime}
\end{align}

\section{Transverse-field Ising chain}
\label{app:7}

For $n=2$, the permutation $V_A^{(2)}$ is the subsystem SWAP operator $\mathbb S_A$. If a weak quench
$H_0\to H_0+\delta\lambda\mathcal O$ is applied to an eigenstate of $H_0$, the $H_0$ contribution produces only a common phase, and Eq.~\eqref{eq:SM-xdot} becomes
\begin{align}
 \dot x_2(0)
 =-2\delta\lambda\left[
 \frac12\langle\{\rho_A,\mathcal O\}\rangle_0
 -\operatorname{Tr}(\rho_A^2)\langle\mathcal O\rangle_0
 \right].
 \label{eq:SM-manybody}
\end{align}
Equivalently, defining $\Delta X\equiv X-\langle X\rangle_{\Psi_0^{\otimes2}}$,
\begin{align}
 \dot x_2(0)
 =-\frac{\delta\lambda}{2}
 \left\langle
 \left\{\Delta\mathbb S_A,
 \Delta[\mathcal O^{(1)}+\mathcal O^{(2)}]\right\}
 \right\rangle_{\Psi_0^{\otimes2}}.
 \label{eq:SM-swap-response}
\end{align}
This form involves only a two-copy SWAP covariance and does not require constructing the modular Hamiltonian.

We now evaluate this response in the open transverse-field Ising chain
\begin{align}
 &H_0(h)=-J\sum_{j=1}^{L-1}\sigma_j^z\sigma_{j+1}^z
 -h\sum_{j=1}^{L}\sigma_j^x,
 \nonumber\\
 &\mathcal O=\partial_hH_0=-\sum_{j=1}^{L}\sigma_j^x,
 \label{eq:SM-TFIM}
\end{align}
using its even-parity ground state and the boundary-adjacent half chain $A=\{1,\ldots,L/2\},\, L_{A} \equiv L/2$. The model is unitarily equivalent to
$\widetilde H_0=-J\sum_j\sigma_j^x\sigma_{j+1}^x-h\sum_j\sigma_j^z$ and
$\widetilde{\mathcal O}=-\sum_j\sigma_j^z$. After the Jordan--Wigner transformation, $\widetilde H_0$ is a quadratic Majorana Hamiltonian, while
$\widetilde{\mathcal O}=-L+2N$ is diagonal in the fermion number $N$ \cite{Pfeuty1970}. This permits calculations for chains much longer than those accessible to spin-basis exact diagonalization.

Let
\begin{align}
 \bm C=(c_1,\ldots,c_L,c_1^\dagger,\ldots,c_L^\dagger)^T,
 \qquad
 \mathcal R=\langle\bm C\bm C^\dagger\rangle_0.
\end{align}
For a pure Gaussian ground state, $\mathcal R$ is a rank-$L$ projector. We write $\mathcal R=QQ^\dagger$, where the columns of the $2L\times L$ matrix $Q$ form an orthonormal basis of quasiparticle annihilators. The restriction to the Nambu components supported in $A$ is denoted by $\mathcal R_A$, and $\mathbb I_{2L_A}$ denotes the identity on this restricted Nambu space. The second-R\'enyi purity follows directly from the restricted correlation matrix \cite{Peschel2003,PeschelEisler2009},
\begin{align}
 P_2=\operatorname{Tr}\rho_A^2
 =\sqrt{\det\!\left[\mathcal R_A^2+(\mathbb I_{2L_A}-\mathcal R_A)^2\right]}.
 \label{eq:SM-gaussian-purity}
\end{align}

The Wick-rotated identity in Eq.~\eqref{eq:SM-imagtime} turns the real-time response into a derivative of this ordinary purity. Introduce the normalized filtered state
\begin{align}
 |\Psi_\lambda\rangle
 =\frac{e^{-\lambda\widetilde{\mathcal O}}|\Psi_0\rangle}
 {\sqrt{\langle\Psi_0|e^{-2\lambda\widetilde{\mathcal O}}|\Psi_0\rangle}}.
 \label{eq:SM-filtered-state}
\end{align}
Since $e^{-\lambda\widetilde{\mathcal O}}$ is proportional to $e^{-2\lambda N}$, the annihilator subspace transforms as
\begin{align}
 &Q_\lambda=D_\lambda Q
 \left(Q^\dagger D_\lambda^2Q\right)^{-1/2},
 \quad
 D_\lambda=e^{\lambda\mathcal K},
 \nonumber\\
 &\mathcal K=\begin{pmatrix}2\bm 1_L&0\\0&-2\bm 1_L\end{pmatrix}.
 \label{eq:SM-Q-filter}
\end{align}
Here $\bm 1_L$ denotes the $L\times L$ identity matrix. Thus $\mathcal R(\lambda)=Q_\lambda Q_\lambda^\dagger$ obeys
\begin{align}
 \mathcal R'(0)=\mathcal K\mathcal R+\mathcal R\mathcal K
 -2\mathcal R\mathcal K\mathcal R.
 \label{eq:SM-R-prime}
\end{align}
Let $\mathcal R_A'\equiv\left.\partial_\lambda\mathcal R_A(\lambda)\right|_{\lambda=0}$ denote the restriction of $\mathcal R'(0)$ to $A$. Writing $\mathcal M_A=\mathcal R_A^2+(\mathbb I_{2L_A}-\mathcal R_A)^2$, differentiation of Eq.~\eqref{eq:SM-gaussian-purity} gives
\begin{align}
 \left.\partial_\lambda\log P_2(\lambda)\right|_0
 =\frac12\operatorname{Tr}\bigl\{\mathcal M_A^{-1}
 \bigl[2(\mathcal R_A\mathcal R_A'
 +\mathcal R_A'\mathcal R_A)-2\mathcal R_A'\bigr]\bigr\}.
 \label{eq:SM-purity-derivative}
\end{align}
For the field quench $h\to h+\delta h$, define
\begin{align}
 \chi_{\theta,2}(h,L)=\frac{\dot\theta_2(0)}{\delta h},
 \qquad
 \chi_{\mathcal D,2}(h,L)
 =\frac{|\dot x_2(0)|}{|\delta h|}.
 \label{eq:SM-susceptibilities}
\end{align}
The short-time identity then yields
\begin{align}
& \chi_{\theta,2}=\frac12\left.\partial_\lambda\log P_2(\lambda)\right|_0,
 \nonumber\\
 &\chi_{\mathcal D,2}=P_2|\chi_{\theta,2}|
 =\frac12\left|\left.\partial_\lambda P_2(\lambda)\right|_0\right|.
 \label{eq:SM-Gaussian-response}
\end{align}
The first quantity isolates phase generation, whereas the second is the initial growth rate of the experimentally available trace distance. They differ by precisely the replica visibility, $\chi_{\mathcal D,2}=\mathcal V_2(0)|\chi_{\theta,2}|$ with $\mathcal V_2(0)=P_2$.

\begin{figure}[t]
 \centering
 \includegraphics[width=0.50\textwidth]{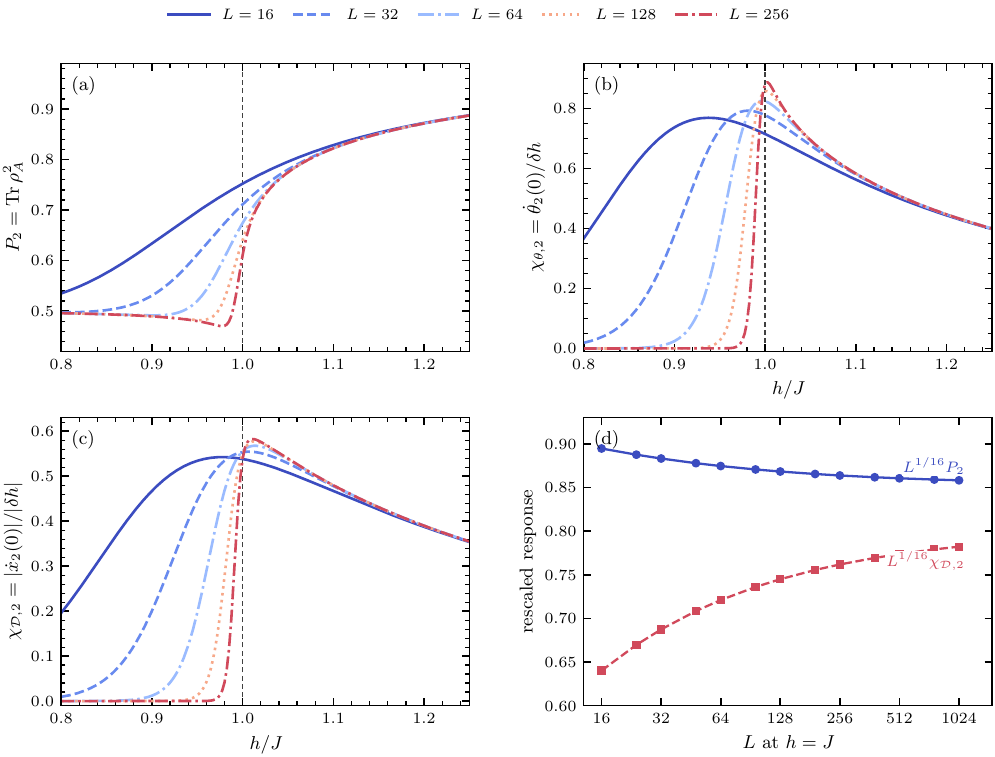}
 \caption{
 Many-body response in the open transverse-field Ising chain for the half-chain subsystem $L_A=L/2$ and $J=1$.
 (a) Initial second-R\'enyi visibility $P_2$.
 (b) Phase susceptibility $\chi_{\theta,2}=\dot\theta_2(0)/\delta h$.
 (c) Operational discrimination susceptibility $\chi_{\mathcal D,2}=|\dot x_2(0)|/|\delta h|=P_2|\chi_{\theta,2}|$.
 The vertical dashed lines mark the thermodynamic critical field $h_c=J$.
 (d) Critical-size dependence. Boundary conformal field theory predicts $P_2\propto L^{-1/16}$ for the critical Ising half chain. The rescaled visibility $L^{1/16}P_2$ and discrimination response $L^{1/16}\chi_{\mathcal D,2}$ remain of order unity and vary slowly over the displayed range.}
 \label{fig:SM-Ising}
\end{figure}

Fig.~\ref{fig:SM-Ising} shows the results. The phase susceptibility develops a progressively narrower maximum near $h/J=1$. For $L=16,32,64,128,256$, its peak is located at $h/J=0.9375,0.9800,0.9975,1.0025,1.0025$, respectively, as shown in Fig.~\ref{fig:SM-Ising}(b). The maximum of the directly measurable susceptibility is also drawn toward the critical region, with locations $0.9775,1.0075,1.0150,1.0125,1.0100$ as shown in Fig.~\ref{fig:SM-Ising}(c). The two maxima need not coincide at finite size: $\chi_{\theta,2}$ measures phase production, whereas $\chi_{\mathcal D,2}$ also contains the visibility $P_2$.

At the critical point, boundary conformal field theory gives for an interval adjacent to an open boundary
\begin{align}
 S_{A,n}^{\mathrm E}=\frac{c_{\rm CFT}}{12}\left(1+\frac1n\right)\log L+O(1)
 \label{eq:SM-boundary-CFT}
\end{align}
for fixed $L_A/L$ \cite{Calabrese2004,IgloiLin2008}, where $c_{\rm CFT}$ is the central charge. With $c_{\rm CFT}=1/2$ and $n=2$, this implies
\begin{align}
 P_2=e^{-S_{A,2}^{\mathrm E}}\propto L^{-1/16}.
 \label{eq:SM-P2-scaling}
\end{align}
Our data up to $L=1024$ show that $\chi_{\theta,2}(J,L)$ remains of order unity and increases slowly from $0.7154$ at $L=16$ to $0.9114$ at $L=1024$. Consequently, the operational response is only weakly suppressed: $\chi_{\mathcal D,2}(J,L)$ changes from $0.5382$ to $0.5072$ over the same range. Panel (d) in Fig.~\ref{fig:SM-Ising} shows that $L^{1/16}P_2$ and $L^{1/16}\chi_{\mathcal D,2}$ remain of order unity and vary only slowly over the available sizes. This behavior is consistent with weak algebraic suppression by the replica visibility, but the present data do not establish an asymptotic scaling law for $\chi_{\mathcal D,2}$. We likewise do not infer an independent critical exponent for $\chi_{\theta,2}$; a field-theoretic analysis of its scaling function is left for future work.

The scaling issue depends on the protocol. The present calculation concerns the initial response, for which $\mathcal V_2(0)=P_2$; it therefore directly tests the visibility entering the short-time discrimination rate. At finite times, a global quench or an extensive subsystem may produce additional overlap suppression, and the exact relation $\mathcal D_2=\mathcal V_2|\sin\theta_2|$ then transfers that loss of contrast directly to the measured distinguishability. By contrast, for a fixed finite subsystem and a local quench, locality limits the region that can affect the measurement at finite time, so enlarging the system beyond the relevant propagation region need not cause an additional extensive suppression \cite{LiebRobinson1972}. Thus severe visibility loss is a protocol-dependent limitation rather than a failure of the operational identity itself.

As a check, we compared Eqs.~\eqref{eq:SM-gaussian-purity}--\eqref{eq:SM-Gaussian-response} with spin-basis exact diagonalization for $L=4,6,8,10$ at $h/J=0.85,1,1.15$. The purity, phase susceptibility, and discrimination susceptibility agree within $3\times10^{-13}$ in all cases.

\section{Relative arrow information and dephasing}
\label{app:6}

For density matrices $\rho$ and $\sigma$, the quantum
relative entropy is defined by
\begin{align}
 D(\rho\|\sigma)
 \equiv
 \operatorname{Tr}\!\left[
 \rho(\log\rho-\log\sigma)
 \right].
\end{align}
To evaluate it for the ancilla states, consider a qubit density
matrix
\begin{align}
 \rho(\boldsymbol r)
 =
 \frac{1}{2}
 \left(
 \mathbb I+\boldsymbol r\cdot\boldsymbol\sigma
 \right),
 \qquad
 |\boldsymbol r|\equiv\mathcal V<1.
\end{align}
Its logarithm can be written as
\begin{align}
 \log\rho(\boldsymbol r)
 =
 \frac{1}{2}
 \log\!\left(\frac{1-\mathcal V^2}{4}\right)\mathbb I
 +
 \operatorname{artanh}(\mathcal V)\,
 \hat{\boldsymbol r}\cdot\boldsymbol\sigma,
 \quad
 \hat{\boldsymbol r}
 \equiv
 \frac{\boldsymbol r}{\mathcal V}.
\end{align}
For the two ancilla states in Eq.~\eqref{eq:SM-rhoF}, the corresponding
Bloch vectors are
\begin{align}
 \boldsymbol r_F
 =
 \mathcal V_n
 \bigl(\cos\theta_n,\sin\theta_n,0\bigr),
 \quad
 \boldsymbol r_B
 =
 \mathcal V_n
 \bigl(\cos\theta_n,-\sin\theta_n,0\bigr).
\end{align}
They have the same length and satisfy
\begin{align}
 \hat{\boldsymbol r}_F\cdot\hat{\boldsymbol r}_B
 =
 \cos(2\theta_n).
\end{align}
Substituting these expressions into the definition of the
relative entropy gives
\begin{align}
 D\!\left(
 \rho_{\mathrm a,F}^{(n)}
 \big\|
 \rho_{\mathrm a,B}^{(n)}
 \right)
 &=
 \mathcal V_n
 \operatorname{artanh}(\mathcal V_n)
 \left(
 1-
 \hat{\boldsymbol r}_F\cdot\hat{\boldsymbol r}_B
 \right)
 \\
 &=
 2\mathcal V_n\sin^2\theta_n\,
 \operatorname{artanh}(\mathcal V_n).
\end{align}
The same expression is obtained after exchanging $F$ and
$B$, because the two Bloch vectors have the same length.
This symmetry is special to the present mirror-related pair;
quantum relative entropy is not symmetric in general. We
therefore define
\begin{align}
 &I_n^{\mathrm{arrow}}
 \equiv
 D\!\left(
 \rho_{\mathrm a,F}^{(n)}
 \big\|
 \rho_{\mathrm a,B}^{(n)}
 \right)
 =
 2\mathcal V_n\sin^2\theta_n\,
 \operatorname{artanh}(\mathcal V_n),
 \nonumber\\
 &0\leq\mathcal V_n<1.
 \label{eq:SM-Iarrow}
\end{align}
The restriction $\mathcal V_n<1$ ensures that both ancilla
states are full rank. The quantity $I_n^{\mathrm{arrow}}$
refers to the full quantum ancilla pair before any measurement
is performed.

The optimal ancilla $Y$ measurement maps the two quantum
states to the classical outcome distributions $p_F(y)$ and
$p_B(y)$ defined in Sec.~\ref{sec:Rep}. Their Kullback--Leibler
divergence is
\begin{align}
 D_{\mathrm{KL}}(p_F\|p_B)
 =
 x_n\log\!\left(\frac{1+x_n}{1-x_n}\right)
 =
 2\mathcal D_n\operatorname{artanh}(\mathcal D_n),
\end{align}
where $x_n=\mathcal V_n\sin\theta_n$ and
$\mathcal D_n=|x_n|$. For $|\theta_n|\ll1$, this becomes
\begin{align}
 D_{\mathrm{KL}}(p_F\|p_B)
 &=
 2\mathcal V_n^2\theta_n^2
 +O(\theta_n^4)
 \nonumber\\
 &=
 2\mathcal V_n^2(1-n)^2
 \left(
 \operatorname{Im}S_{A,n}^F
 \right)^2
 +O(\theta_n^4).
 \label{eq:SM-KL-expansion}
\end{align}
Thus $I_n^{\mathrm{arrow}}$ is the quantum relative entropy
of the two ancilla states, whereas
$D_{\mathrm{KL}}(p_F\|p_B)$ is the relative entropy retained
after the specific $Y$ measurement.

Because measurement cannot increase quantum relative entropy, and by the classical Pinsker bound,
\begin{align}
 I_n^{\mathrm{arrow}}
 \geq D_{\mathrm{KL}}(p_F\|p_B)
 \geq 2\mathcal D_n^2.
 \label{eq:SM-hierarchy}
\end{align}
The first inequality can be strict even though the $Y$ measurement is Helstrom optimal: minimum single-shot error and preservation of the full relative entropy are different statistical tasks \cite{Kullback1951,Fedotov2003}.

Consider the phase-damping channel
\begin{align}
 \sigma_x\mapsto\eta\sigma_x,
 \qquad
 \sigma_y\mapsto\eta\sigma_y,
 \qquad
 \sigma_z\mapsto\sigma_z,
 \qquad 0\leq\eta\leq1.
\end{align}
The visibility becomes $\eta\mathcal V_n$, while the phase is unchanged. The retained and lost arrow information are
\begin{align}
 I_{n,\eta}^{\mathrm{arrow}}
 =2\eta\mathcal V_n\sin^2\theta_n
 \operatorname{artanh}(\eta\mathcal V_n),
\end{align}
\begin{align}
 \Delta_\eta I_n^{\mathrm{arrow}}
 &=I_n^{\mathrm{arrow}}-I_{n,\eta}^{\mathrm{arrow}}\nonumber\\
 &=2\mathcal V_n\sin^2\theta_n\left[
 \operatorname{artanh}\mathcal V_n
 -\eta\operatorname{artanh}(\eta\mathcal V_n)
 \right]\geq0.
 \label{eq:SM-dephasing-loss}
\end{align}
Panel (c) of Fig.~\ref{fig:twoqubit} in the main text uses $p=p_\star$, $gt=0.45$, and Eq.~\eqref{eq:SM-dephasing-loss}.

More generally, monotonicity of quantum relative entropy under a completely positive trace-preserving map $\Lambda$ \cite{Lindblad1975,Uhlmann1977} gives
\begin{align}
 D(\rho_{\mathrm a,F}^{(n)}\|\rho_{\mathrm a,B}^{(n)})
 \geq D\!\left(\Lambda(\rho_{\mathrm a,F}^{(n)})
 \middle\|\Lambda(\rho_{\mathrm a,B}^{(n)})\right).
\end{align}
To state the equality condition explicitly, set
\(\sigma_n=\rho_{\mathrm a,B}^{(n)}\).  The Petz recovery map associated with
\(\sigma_n\) and \(\Lambda\) is
\begin{align}
 \mathcal R_{\sigma_n,\Lambda}(X)
 =\sigma_n^{1/2}\Lambda^\dagger\!\left[
 \Lambda(\sigma_n)^{-1/2}X\Lambda(\sigma_n)^{-1/2}
 \right]\sigma_n^{1/2},
 \label{eq:SM-petz}
\end{align}
where $\Lambda^\dagger$ denotes the Hilbert--Schmidt adjoint of $\Lambda$, and the inverses are restricted to the relevant supports. Equality in data processing holds if and only if
\begin{align}
 \mathcal R_{\sigma_n,\Lambda}
 \!\left[\Lambda(\rho_{\mathrm a,F}^{(n)})\right]
 =\rho_{\mathrm a,F}^{(n)},
\end{align}
while the reference state \(\sigma_n\) is recovered by construction \cite{Petz1986}.

\section{Schmidt-diagonal class}
\label{app:4}

Consider the Schmidt-diagonal form
\begin{align}
 |\Psi_0\rangle=\sum_j\sqrt{p_j}|j\rangle_A|j\rangle_{\bar A},
 \qquad
 H|j,j\rangle=E_j|j,j\rangle.
\end{align}
Then \cite{Misumi2026}
\begin{align}
 \chi(t)=\sum_jp_je^{-iE_jt},
 \qquad
 G_n^F(t)=\sum_jp_j^ne^{-inE_jt},
\end{align}
and
\begin{align}
 Z_n^F(t)=\frac{\sum_jp_j^ne^{-inE_jt}}
 {(\sum_jp_je^{-iE_jt})^n}.
 \label{eq:SM-Schmidt-Z}
\end{align}
Defining normalized weights
\begin{align}
 w_j^{(n)}=\frac{p_j^n}{\sum_kp_k^n},
\end{align}
the initial imaginary response is
\begin{align}
 \left.\partial_t\operatorname{Im}S_{A,n}^F\right|_0
 =-\frac{n}{1-n}\left(\sum_jw_j^{(n)}E_j-\sum_jp_jE_j\right).
\end{align}
The $n\to1$ limit is $-\operatorname{Cov}_p(-\log p_j,E_j)$, where
\begin{equation*}
 \operatorname{Cov}_p(a,b)
 \equiv\sum_jp_ja_jb_j
 -\left(\sum_jp_ja_j\right)\left(\sum_jp_jb_j\right).
\end{equation*}

\section{Complete two-qubit formulas}
\label{app:5}

Let
\begin{align}
 |\Psi_0\rangle=\sqrt p|00\rangle+e^{i\phi}\sqrt q|11\rangle,
 \qquad q=1-p,
\end{align}
with
\begin{align}
 H=J\sigma_z\otimes\sigma_z+g\sigma_x\otimes\sigma_x.
\end{align}
This model was used in Ref.~\cite{Misumi2026} to illustrate the modular-covariance response. The replica-interferometric formulas below are derived here. Writing $r=\sqrt{pq}$, $c=\cos(gt)$, and $s=\sin(gt)$,
\begin{align}
 \chi(t)=e^{-iJt}\left(c-2ir\cos\phi\,s\right).
\end{align}
The reduced transition matrix is diagonal,
\begin{align}
 T_A^F(t)=e^{-iJt}
 \begin{pmatrix}
 pc-ire^{i\phi}s&0\\
 0&qc-ire^{-i\phi}s
 \end{pmatrix}.
\end{align}
For $n=2$, define
\begin{align}
 U(t)&=c^2(p^2+q^2)+2rcs(p-q)\sin\phi
 -2r^2s^2\cos(2\phi),\\
 W(t)&=2rcs\cos\phi.
\end{align}
Then
\begin{align}
 G_2^F(t)=e^{-2iJt}[U(t)-iW(t)],
 \quad
 \mathcal V_2(t)=\sqrt{U(t)^2+W(t)^2}.
\end{align}
The calibrated signed signal is
\begin{align}
 x_2(t)
 &=\frac{\operatorname{Im}[G_2^F(t)\chi(t)^{*2}]}{|\chi(t)|^2}
 \nonumber\\
 &=\frac{2rcs\cos\phi\,[c(p-q)+2rs\sin\phi]^2}
 {c^2+4r^2s^2\cos^2\phi}.
 \label{eq:SM-x2-general}
\end{align}
For $\phi=0$ and $\delta=p-q=2p-1$,
\begin{align}
 x_2(t)=\frac{2r\delta^2c^3s}{1-\delta^2s^2},
 \qquad
 x_2(t)=2gr\delta^2t+O(t^3).
\end{align}
The short-time magnitude is maximized by $\delta^2=2/3$, or
\begin{align}
 p=p_\star,\qquad 1-p_\star,
 \qquad
 p_\star=\frac12\left(1-\sqrt{\frac23}\right).
\end{align}

\bibliography{references}

\end{document}